%% file: dino-cxr.tex
\documentclass{article}

\usepackage{arxiv}

\usepackage[utf8]{inputenc} 
\usepackage[T1]{fontenc}    
\usepackage{hyperref}       
\usepackage{booktabs}       
\usepackage{nicefrac}       
\usepackage{microtype}      
\usepackage{lipsum}
\usepackage{graphicx}
\usepackage{float}
\usepackage[table]{xcolor}
\usepackage{multirow}
\usepackage{makecell}
\usepackage{longtable}
\usepackage{cite}
\usepackage{amsmath,amssymb,amsfonts}
\usepackage{graphicx}
\usepackage{textcomp}
\usepackage{xcolor}
\usepackage{longtable}
\usepackage{booktabs}
\usepackage{subcaption}
\usepackage{url}
\usepackage{tabularx}
\usepackage{array}
\usepackage{multirow}

\definecolor{LightCyan}{rgb}{0.88,1,1}

\title{DINO-CXR: A Self Supervised Method Based on Vision Transformer for Chest X-Ray Classification}

\author{
  Mohammadreza Shakouri \\
  Department of Computer Engineering\\
  Shahid Bahonar University of Kerman \\
  Kerman, Iran \\
  \texttt{mohammadrezashakouri@eng.uk.ac.ir} \\
  \And
  Fatemeh Iranmanesh \\
  Department of Computer Engineering\\
  Shahid Bahonar University of Kerman \\
  Kerman, Iran \\
  \texttt{firanmanesh@eng.uk.ac.ir} \\
   \And
  Mahdi Eftekhari \\
  Department of Computer Engineering\\
  Shahid Bahonar University of Kerman \\
  Kerman, Iran \\
  \texttt{m.eftekhari@uk.ac.ir} \\
}
\date{May 1, 2023}

\begin{document}
\maketitle

\begin{abstract}
The limited availability of labeled chest X-ray datasets is a significant bottleneck in the development of medical imaging methods. Self-supervised learning (SSL) can mitigate this problem by training models on unlabeled data. Furthermore, self-supervised pretraining has yielded promising results in visual recognition of natural images but has not been given much consideration in medical image analysis. In this work, we propose a self-supervised method, DINO-CXR, which is a novel adaptation of a self-supervised method, DINO, based on a vision transformer for chest X-ray classification. A comparative analysis is performed to show the effectiveness of the proposed method for both pneumonia and COVID-19 detection. Through a quantitative analysis, it is also shown that the proposed method outperforms state-of-the-art methods in terms of accuracy and achieves comparable results in terms of AUC and F-1 score while requiring significantly less labeled data.
\end{abstract}


\keywords{Deep Learning \and Self-supervised Learning \and Chest X-ray Classification
}

\input{introduction}

\input{background}

\input{related-work}

\input{methods}

\input{Experiment-Setup}

\input{analysis-results}

\input{conclusion}

\Urlmuskip=0mu plus 1mu\relax
\bibliographystyle{acm}
\bibliography{dino-cxr}

\end{document}

%% file: introduction.tex
\section{Introduction}

The use of medical image classification has grown significantly in the last decade~\cite{huang2020unet}. Despite a recent slowdown~\cite{hong2020trends}, medical imaging remains a vital diagnostic tool, particularly in the detection of pneumonia and COVID-19. Pneumonia is the leading cause of death in children under the age of five worldwide~\cite{owayed2000underlying}. Additionally, the outbreak of the novel COVID-19 in late 2019 rapidly evolved into a global pandemic, affecting millions of people worldwide. Chest X-rays are one of the most common ways to diagnose pneumonia and COVID-19. They can show lung abnormalities that are consistent with pneumonia or COVID-19.
\par
Chest X-ray (CXR) classification is a more challenging task than natural image classification in that, 1) the identification of diseases in chest X-rays may be contingent on the presence of abnormalities in a small number of pixels, 2) chest X-rays are different from natural images in terms of their data attributes: X-rays are larger, grayscale, and have similar spatial structures across images, 3) the number of unlabeled chest X-ray images is significantly less than the number of unlabeled natural images. \par

Acquiring knowledge from a small amount of labeled data is a significant challenge in machine learning, especially in medical image analysis, where the annotation of medical images is a labor-intensive and costly process that requires the involvement of specialists. In recent years,  
self-supervised learning, particularly contrastive learning~\cite{hadsell2006dimensionality}, has emerged as a promising new approach for addressing the challenges of limited labeled data in various domains by creating pre-trained models from unlabeled data for subsequent fine-tuning on labeled data. Contrastive learning frameworks such as DINO~\cite{caron2021emerging} maximize agreement between positive image pairs using a contrastive loss function, while differing in their data augmentation and sampling strategies. 

There are two widely used methods for pretraining models to learn from a limited amount of labeled data. The first method is~\textit{supervised pretraining}, which involves training the model on a large labeled dataset (e.g., ImageNet). The second method is~\textit{self-supervised pretraining}, which involves using contrastive learning on unlabeled data~\cite{chen2020big}.

\par
Transformers~\cite{vaswani2017attention} have recently emerged as a viable alternative to convolutional neural networks (CNNs) for visual recognition tasks. The success of transformers has inspired many following works that apply them to various computer vision tasks~\cite{dosovitskiy2020vit,zheng2021rethinking}. 

\begin{figure}[t]
     \centering
     \includegraphics[width=0.6\textwidth]{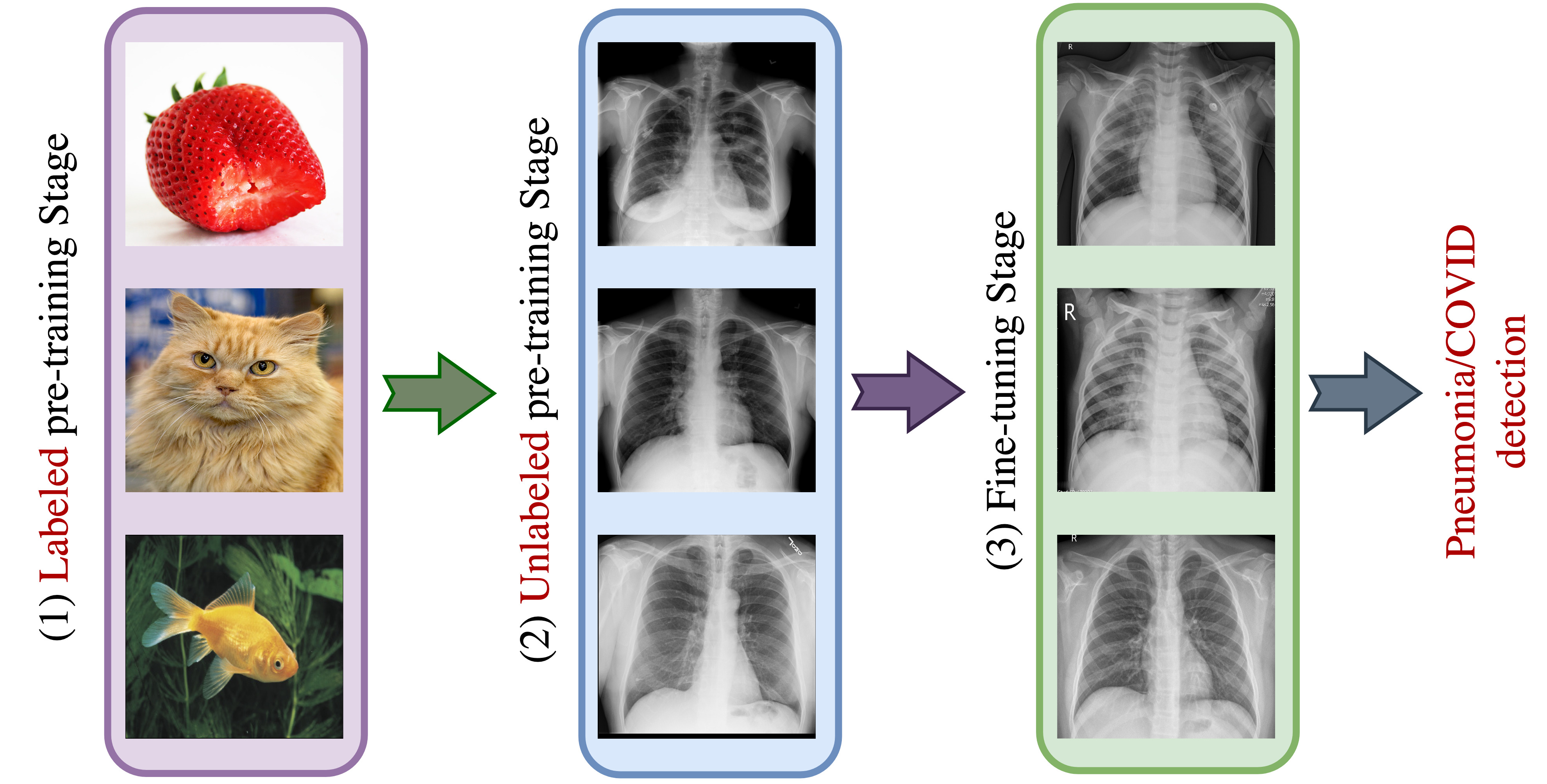}
     \caption{The pre-training strategy for self-supervised models.}
     \label{fig:scheme pre-training}
\end{figure}

This paper presents DINO-CXR, an innovative adaptation of a self-supervised technique, DINO, based on ViTAEv2~\cite{xu2021vitae} vision transformer as the backbone for CXR classification. ViTAEv2 is a promising new vision transformer architecture that takes advantage of inductive biases such as \textit{locality} and \textit{scale-invariance}, which we demonstrate that are useful for CXR classification.  This goes beyond current state-of-the-art techniques that rely on CNN or ViT~\cite{dosovitskiy2020image} as a backbone. We also modify DINO to make it less computationally expensive with better performance for CXR classification.

Self-supervised learning is used as a pre-training strategy for CXR classification. In this regard, first, a model is pre-trained using supervised learning on a labeled dataset of natural images (Fig.~\ref{fig:scheme pre-training}). Next, self-supervised pre-training is employed on a large dataset of unlabeled CXR images. Finally, the model is fine-tuned on a small dataset of labeled CXR images.

Various experiments are conducted to show the effectiveness of DINO-CXR for CXR classification. An extensive comparison is performed between self-supervised approaches with different networks and frameworks, and it is demonstrated that DINO-CXR outperforms other methods in terms of accuracy, AUC, and F-1 score. DINO-CXR also outperforms state-of-the-art self-supervised methods for pneumonia and COVID detection in terms of accuracy. In addition, DINO-CXR achieves comparable results for CVOID-19 detection in terms of precision and F-1 score while using significantly less labeled data.

In summary, this paper presents the following contributions:
\begin{itemize}
\item We propose a self-supervised method by adapting DINO based on a vision transformer, ViTAEv2, for chest X-ray classification.
\item We modify DINO to make it less computationally expensive while achieving better results for CXR classification.
\item To the best of our knowledge, this work is the first work to compare self-supervised pre-training approaches with different networks and frameworks for CXR classification.
\item 
The proposed method is shown to outperform state-of-the-art self-supervised methods in terms of accuracy for pneumonia and COVID-19 detection. It is also shown to achieve comparable results for COVID-19 detection in terms of precision and F-1 score while using significantly less (6\%) of labeled data.
\end{itemize}

%% file: background.tex
\section{Background}
This section provides background information on self-supervised frameworks and backbones used in this study.

\subsection{Self-supervised frameworks}
\label{SLL approaches}
\begin{figure}[t]
    \centering
    \includegraphics[width=0.6\textwidth]{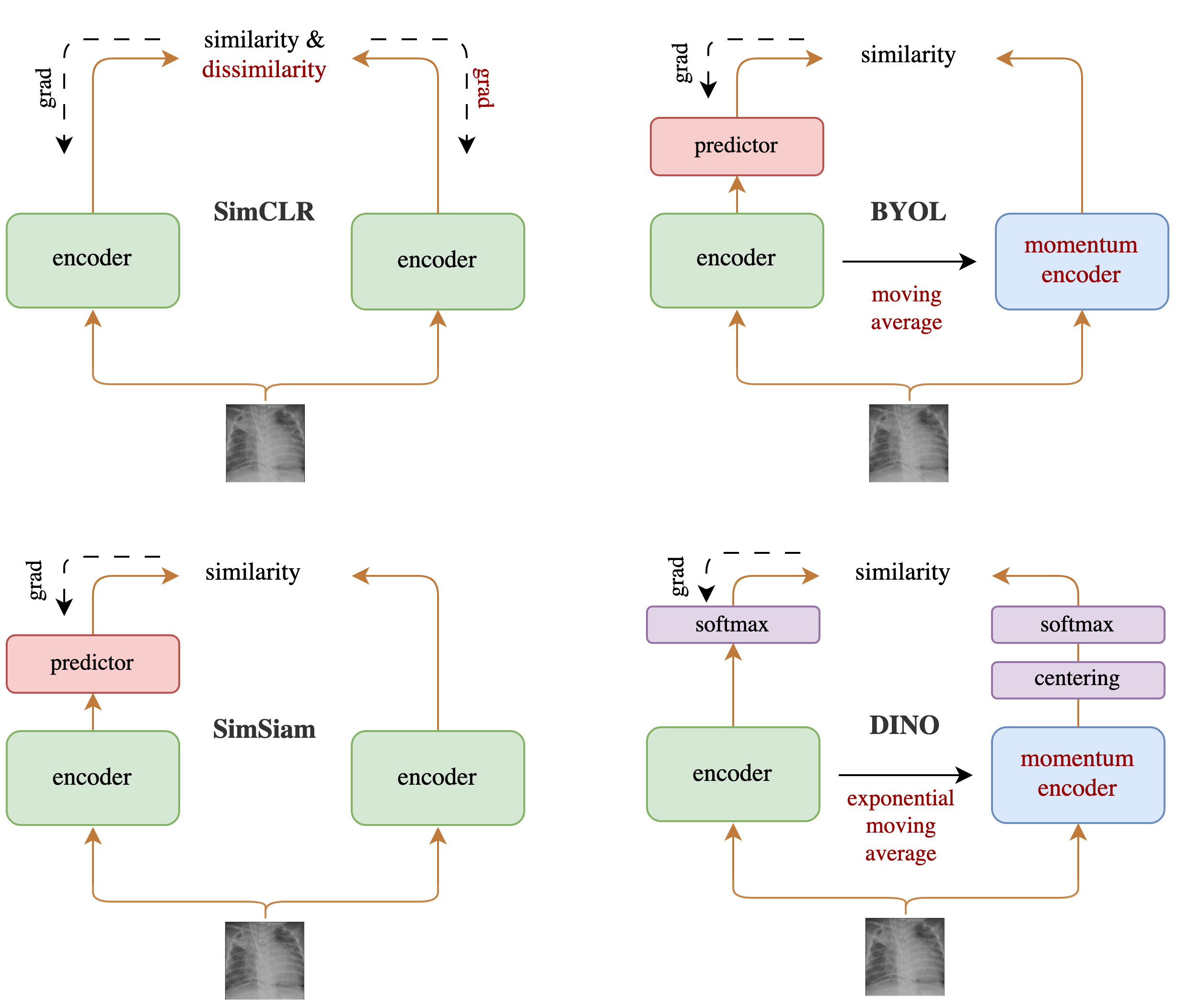}
    \caption{\textbf{Self supervised approaches architectures}. Two \textbf{augmented views} of one image are processed by the encoder network. The encoder includes all layers that can be shared between both branches. The dashed lines represent the gradient propagation flow.}
    \label{fig:SSLArch}
\end{figure}

An overview of the self-supervised frameworks used in this work is provided in this section. Figure~\ref{fig:SSLArch} shows these frameworks.

\subsubsection{SimCLR}SimCLR~\cite{chen2020simple} is a simple framework for contrastive learning of visual representations. It learns representations by maximizing the similarity between differently augmented views of the same data sample and minimizing the similarity to other images via a contrastive loss. Two augmented images are passed through an encoder. Then a nonlinear fully connected layer is applied to get representations. The goal is to maximize the similarity between these two representations. 
SimCLR uses negative samples to avoid finding a collapsing solution. Similarity minimization between each input and its negative pairs is the key to preventing collapse.

\subsubsection{BYOL}BYOL~\cite{grill2020bootstrap} consists of two neural networks: The online and target networks. These two networks are interacted and learn from each other. An augmented view of an image is fed to the online network to predict the target network representation of another variant augmented view of the same image. At the same time, the target network is updated with a slow-moving average of the online network. \par
While SimCLR uses contrastive loss between positive and negative samples, BYOL uses only positive samples in the loss function. Apparently, BYOL seems to be doing self-supervised learning without contrasting different images. However, it appears that the main reason BYOL works is that it does a form of contrastive learning- just through an indirect mechanism.\par

\subsubsection{SimSiam}SimSiam~\cite{chen2021exploring} is a self-distillation method that feeds two different augmented views of one image to two same encoders, which includes a backbone and a projection MLP. The output of one side passed through a prediction MLP, and a stop-gradient technique is used on the other side to prevent collapse. The model aims to maximize the similarity between the output of both sides by using a contrastive loss function. \par
The noticeable point to consider is that SimSiam does not rely on negative sample pairs, momentum encoder, and large batches. With this in mind, SimSiam can be considered a "Variant of BYOL that does not use the momentum encoder" and a "SimCLR without negative sample pairs".

\subsection{Backbones}
In this work, we have modified the backbone architecture of each self-supervised framework mentioned in Section \ref{SLL approaches}, and compared its performance with other existing approaches. We explain the backbone architectures used in this work below.

\subsubsection{ResNet}
ResNet \cite{kaming2016resnet} is a specific type of convolutional neural networks. Before Residual Neural Networks, the Vanishing gradient happened with the increase in number of layers in the network. Thus when we increased number of layers, the training and test error rate increased. In order to solve this problem, residual blocks were introduced and ResNets were made by stacking these residual blocks together.
ResNet-50 is a ResNet neural network with 50 layers. This network uses 1×1 convolutions, known as a “bottleneck” to reduce the number of parameters.

\subsubsection{Vision transformer}
\label{vision transformer}
Vision Transformer(ViT) \cite{dosovitskiy2020vit} is an architecture that trains the encoder part of the original transformer on ImageNet, achieving very good results compared to convolutional architectures. The input to the encoder is a sequence of embedded image patches including a learnable class embedding, which is also augmented with positional information. A classification head attached to the output of the encoder receives the value of the learnable class embedding to generate a classification output.

\subsection{Transfer learning}
Despite recent advances in deep learning, transferring knowledge from natural images to medical image analysis is a widely used approach \cite{liu2020deep,azizi2021big,menegola2017knowledge}. In this regard, empirical studies show that this improves performance \cite{alzubaidi2020towards}. However, Raghu et al.\cite{raghu2019transfusion} investigated the efficacy of ImageNet pretraining. They found that transfer learning from ImageNet can speed up convergence but does not always improve performance, especially in medical imaging contexts.

%% file: related-work.tex
\section{Related Work}
In recent years, there has been a significant advancement in the use of deep learning techniques to detect pneumonia and COVID-19 using medical imaging data, particularly chest X-rays. These approaches can be broadly classified into supervised and self-supervised learning methods.

\subsection{Supervised CXR Classification}
Supervised learning techniques have been broadly utilized for pneumonia and COVID-19 detection by utilizing big annotated datasets to train deep-learning models. In recent years, many studies have investigated the use of deep neural networks for CXR classification tasks \cite{wang2020covid,afshar2020covid}.
Kermany et al.\cite{kermany2018identifying} describe a deep learning framework for the screening of patients with common treatable blinding retinal diseases. The framework uses transfer learning, which trains a neural network with a fraction of the data of conventional approaches.
 
In the context of COVID-19 detection, Wang et al.~\cite{wang2020covid} used a supervised learning approach to train a deep learning model for COVID-19 detection using chest X-ray images.

\subsection{Self-supervised CXR Classification }
Self-supervised learning, particularly contrastive learning, has gained significant attention in recent years for its ability to learn useful representations from unlabeled data. In the context of pneumonia and COVID-19 detection, leveraging self-supervised learning can be advantageous due to the scarcity of labeled medical images. Recent studies have demonstrated the potential of self-supervised learning for CXR classification~\cite{han2021pneumonia,gazda2021self}. Han et al.~\cite{han2021pneumonia} proposed a pneumonia detection method based on radiomic features and contrastive learning that uses self-supervised learning to extract features from chest X-rays.

In 2020, Gazda et al.~\cite{gazda2021self} proposed a self-supervised deep convolutional neural network for chest X-ray classification, including pneumonia and COVID-19 detection.

%% file: methods.tex
\section{Proposed Method}

In this section, we present our proposed method, DINO-CXR, for chest X-ray classification. DINO-CXR is a novel adaptation of DINO based on a vision transformer, ViTAEv2, for chest X-ray classification.

\subsection{DINO}\label{dino}
DINO (self-\textbf{di}stillation with \textbf{no} labels) is a knowledge distillation-based contrastive self-supervised learning algorithm that maximizes the similarity of representations generated from augmented views of the same input image. This similarity is measured with a cross-entropy loss.

Two augmented images are fed to student and teacher networks with the same architecture but different parameters. Output vectors are normalized by softmax with temperature parameters. Teacher is a momentum network, meaning that its parameters are updated with the exponential moving average of student parameters.
DINO avoids collapse by centering and sharpening the momentum of teacher outputs. This balances their effects and prevents collapse. 

\subsection{ViTAE}
\label{vitaev2}
ViTAE is a vision transformer model which incorporates inherent inductive biases, such as \textit{locality} and \textit{scale-invariance}. Inductive bias refers to a set of assumptions or biases that help machine learning models to achieve the power of generalization. In other terms, inductive bias enables the algorithm to prioritize one solution or interpretation over another, regardless of the observed data~\cite{mitchell1980need}.

ViTAE employs two basic cells: reduction cell (RC) and normal cell (NC). The reduction cell embeds the input images into tokens with multi-scale context and local information. RC has two parallel branches that model locality and long-range dependency, respectively. These branches are followed by a feed-forward neural network for feature transformation. One of the branches includes a Pyramid Reduction module to extract multi-scale context and a Multi-Head Self-Attention module to model long-range dependencies, and the other contains a Parallel Convolutional Module to embed local context into the tokens.
The normal cell is utilized to enhance the modeling of both local and long-range dependencies within the tokens. Technically, NC has a similar structure to the reduction cell but does not include the Pyramid Reduction module.

ViTAEv2~\cite{zhang2023vitaev2}, a new version of ViTAE, used another inductive bias, such as local window attention introduced in~\cite{liu2021swin}, in the RC and NC modules. As a result, the model achieves a better balance between memory usage, speed, and performance.

\begin{figure*}[!ht]
    \centering
    \includegraphics[width=0.8\textwidth]{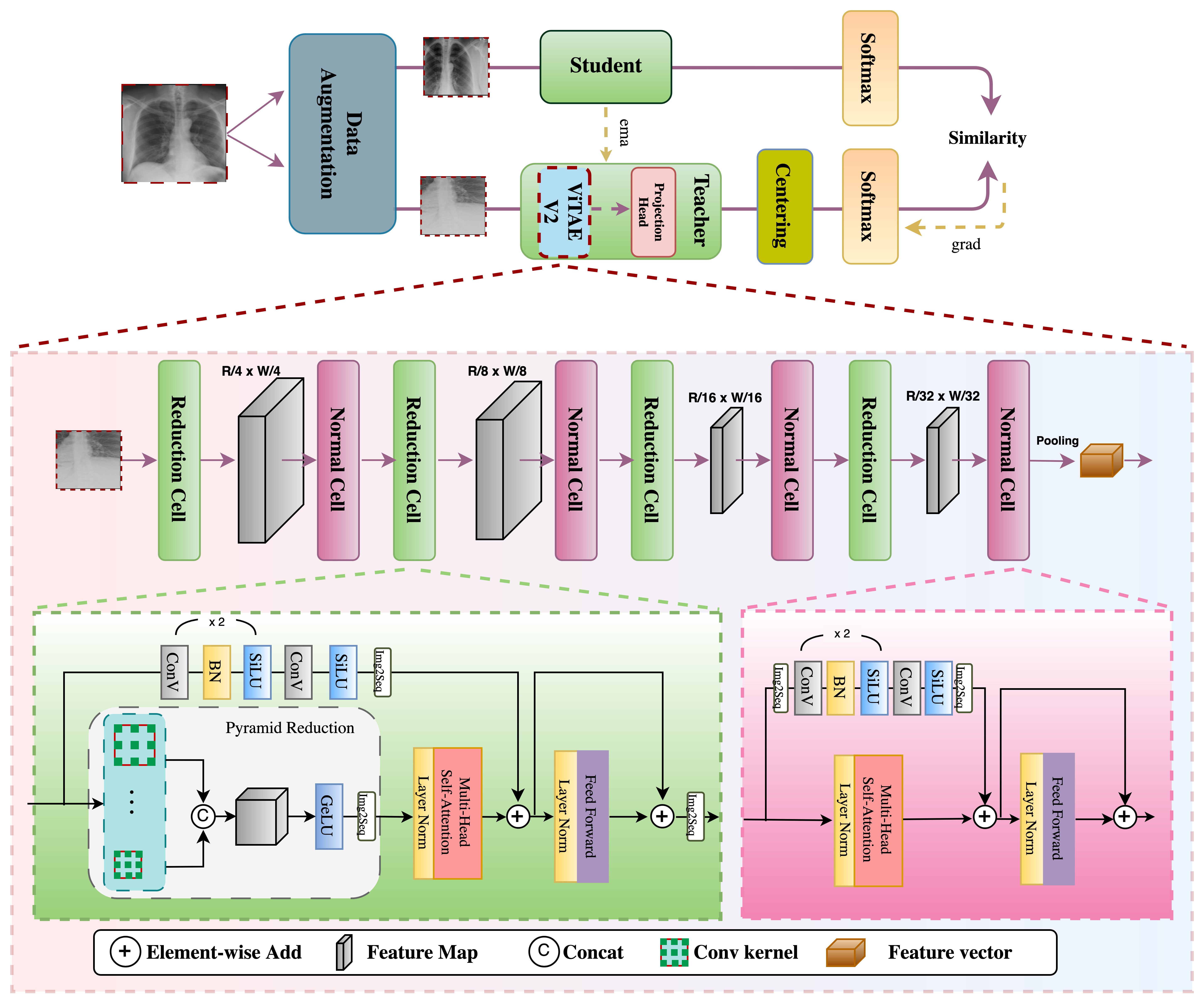}
    \caption{The architecture of DINO-CXR}
    \label{fig: proposed approach}
\end{figure*}
\subsection{DINO-CXR}
Vision transformers have demonstrated considerable promise in various computer vision tasks due to their robust ability to model long-range dependencies using the self-attention mechanism. Long-range dependency refers to the ability to identify patterns that are spread out over the entire image. This is important for pneumonia/COVID-19 detection, as the disease can often manifest as a combination of different abnormalities in the lungs. 

On the other hand, current methods lack inductive biases such as \textit{locality} and \textit{scale-invariance}, and requires large-scale training datasets and longer training schedules to achieve optimal performance.

\textit{Locality} in images refers to the spatial relationships between nearby pixels or regions. In the context of chest X-rays, these structures can help identify patterns and features that are indicative of pneumonia, such as the shape and texture of the lungs, the presence of consolidations, or the appearance of ground-glass opacities. By capturing these local structures, a model can be trained to differentiate between healthy and pneumonia-affected lungs.

Moreover, \textit{scale invariance} refers to the ability of a model to recognize patterns and structures at different scales or sizes within an image. In the field of chest X-rays, scale invariance is important because the size and shape of the lungs, as well as the appearance of pathological features, can vary significantly between patients due to factors such as age, body size, and the severity of the condition.

One of the key innovations of DINO is the use of multi-cropping, which involves training the model on multiple crops of the same image. However, multi-cropping also has a high computational cost. This is because the model needs to be trained on a large number of crops, which can significantly increase the training time and memory usage. To address this issue, we replace multi-cropping with our own data augmentation strategy (Section~\ref{Augmentation Strategy}), which uses a fixed size for the teacher and student network inputs. By doing so, our adapted DINO requires less computation and it is demonstrated that the adapted DINO has better performance than the original DINO. 

With all that said, DINO-CXR is built based on ViTAEv2 which is able to model long-range dependency as well as local structures and deal with scale variance. Thus, DINO-CXR is capable of effectively capturing these properties, making it more likely to achieve high performance in detecting pneumonia/COVID-19.

Figure~\ref{fig: proposed approach} shows the architecture of DINO-CXR. The model feeds two randomly transformed versions of an input image to the student and teacher networks, respectively. The student and teacher networks have the same architecture, but they have different sets of parameters. A stop-gradient operator was applied to the teacher to propagate gradients only through the student. The teacher parameters are updated using an exponential moving average of the student parameters. For instance, we depicted the inside of the teacher block, which includes ViTAEv2 as an encoder and projection head.

\subsection{Augmentation Strategy}

\label{Augmentation Strategy}
In this work, some data augmentation inspired by the study of image augmentations for Siamese networks was investigated ~\cite{van2023exploring}.
Random resized cropping is employed to construct crops with a random aspect ratio of 3/4 -- 4/3  and scale parameter of 0.3 -- 0.9; We utilize color distortion that is composed of color jittering and color dropping. Finally, Gaussian blurring is applied to the image with a probability of 50\%. The standard deviation of the Gaussian kernel is randomly sampled from the range [0.1, 2.0], and the kernel size is set to be 10\% of the image height or width.

%% file: Experiment-Setup.tex
\section{Experiments Setup}\label{sec:experimentstp}
In this section, our experiments setup including setups used for pre-training and fine-tuning are described. All experiments were conducted on Google Colab Pro, which provides access to a powerful GPU. The specific GPU used was an NVIDIA A100 with 40GB of RAM.

\subsection{Datasets}
In this study, three CXR datasets are utilized (Table~\ref{tab:datasetsinfo}). Initially, for pre-training, ChestX-ray14-v3 \footnote{\url{https://www.kaggle.com/datasets/haipham1202/chestx-ray14-v3}} is used, which contains 13,000 X-rays images of three classes - Normal, Pneumonia, and Covid-19. Although labels are available, the unlabelled dataset is employed for all self-supervised pre-training experiments.

For the first downstream task, the Cell dataset~\cite{kermany2018identifying} is utilized, which comprises 5,323 chest X-ray images from children. All chest X-ray labels were generated by two expert physicians and validated by a third physician.

For the second downstream task, the COVIDGR dataset \cite{tabik2020covidgr} is used which contains 426 positive (Covid-19) and 426 negative (Non-Covid-19) chest x-rays. It is important to note that 76 of the 426 COVID-19 patients diagnosed positive by PCR had normal chest X-rays, which makes the task of classifying COVID-19 cases more challenging.
\renewcommand{\arraystretch}{1.4}
\begin{table}[!t]
    \centering
    \begin{tabular}{lccccc}
        \Xhline{2\arrayrulewidth}
          Task &  Dataset  &  Samples  &  Negative  &  Positive  \\ \hline
          Pretext&  ChestX-ray14-v3 & 13,000 & - & -  \\ \hline
          \multirow{2}{0.9cm}{Target} & Cell & 5,323 & 1,349 & 3,883\\
          & COVIDGR & 852 & 426 & 426 \\ 
          \Xhline{2\arrayrulewidth}
    \end{tabular}
    \vspace{3px}
    \caption{Datasets used in this study}
    \label{tab:datasetsinfo}
\end{table}

\subsection{Pretraining Protocol}
DINO-CXR is pre-trained with an effective batch size of 64. We use Adam optimizer over 100 epochs with a learning rate of 0.000125 and fixed the weight decay value from 1e-7 to 1e-6.
A summary of all the setups utilized in this study is presented in table~\ref{tab:pre-setup}.

\renewcommand{\arraystretch}{1.3}
\begin{table}[!ht]
    \centering
    \begin{tabular}{lcccc}
        \Xhline{2\arrayrulewidth}
         Framework & Model & Learning rate & Optimizer & Weight decay \\ \hline
         \multirow{3}{2cm}{SimSiam} & ResNet-50 & 0.025 & SGD & 1e-4  \\ 
         & ViT-S/16  & 0.025 & SGD & 1e-4  \\ 
         & ViTAEv2  & 0.0125 & SGD & 1e-4  \\ \hline
         \multirow{3}{2cm}{SimCLR} & ResNet-50 & 0.15 & LARS & 1e-5  \\ 
         & ViT-S/16 & 0.15 & LARS & 1e-5  \\ 
         & ViTAEv2  & 0.075 & LARS & 1e-5  \\ \hline
         \multirow{3}{2cm}{BYOL} & ResNet-50 & 0.1 & LARS & 1e-5  \\ 
         & ViT-S/16  & 0.1 & LARS & 1e-5  \\ 
         & ViTAEv2  & 0.05 & LARS & 1e-5  \\ \hline
         \multirow{3}{2.5cm}{Adapted DINO} & ResNet-50  & 0.00025 & Adam & 1e-6 to 1e-5  \\ 
         & ViT-S/16  & 0.00025 & Adam & 1e-6 to 1e-5  \\ 
         &  ViTAEv2  & 0.000125 & Adam & 1e-7 to 1e-6 \\  \Xhline{2\arrayrulewidth}
    \end{tabular}
    \vspace{5px}
    \caption{All pre-training setups that are used in this study.}
    \label{tab:pre-setup}
\end{table}

\subsection{Fine-tuning Protocol}
We evaluate the learned representations on the Cell and COVIDGR datasets using the standard linear evaluation protocol for all experiments. This protocol involves training a linear classifier on top of the frozen representation without updating the network parameters \cite{kolesnikov2019revisiting}. We employ batch size of 512 for all SSL approaches with CNN and ViT backbones and batch size of 256 for ViTAEv2, SGD optimizer with a momentum parameter of 0.9 over 50 epochs.
We resize images to 256x256 for preprocessing and took a single center crop of 224x224.

A comprehensive search over the hyperparameter space is performed. We select the learning rate and weight decay after a grid search of the learning rate space \{1e-2, 1e-3, 1e-4\} and the weight decay space \{1e-3, 1e-4, 1e-5\}. Additionally, the learning rate is linearly ramped up during the first 10 epochs to prevent the model from overfitting.



%% file: analysis-results.tex
\section{Experiments \& Results}\label{sec:analysis}

In this section, we present results for the ablation study to show the effectiveness of DINO-CXR. Then we present results for pneumonia and COVID-19 detection.
\renewcommand{\arraystretch}{1.4}
\begin{table}[t]
    \centering
    \begin{tabular}{lcccccc}
        \Xhline{2\arrayrulewidth}
          Network & Framework & ACC & AUC & F1-score & GPU RAM \\ \hline
          \rowcolor{LightCyan}
          ResNet-50 & Adapted DINO  & \textbf{94.0685} & \textbf{0.9405} & \textbf{0.9466} & \textbf{25.9} GB\\
          & DINO  & 93.7343 & 0.9381 & 0.9432 & 39 GB \\
         \Xhline{2\arrayrulewidth}

    \end{tabular}
    \vspace{3px}
    \caption{Comparing our adapted DINO with original DINO}
    \label{tab:resultdino}
\end{table}
\subsection{Ablation Study}
In this section, we first evaluate our adapted DINO compared to original DINO for chest X-ray classification. And then compare DINO-CXR with different combinations of networks and frameworks.
Table~\ref{tab:resultdino} illustrates the results of comparing our adapted DINO vs. original DINO. ResNet-50, the most common backbone, is used for this comparison. Adapted DINO outperforms the original DINO in terms of accuracy, AUC, and F1-score while requiring significantly less computational resources. 

Next we compare DINO-CXR with different combinations of networks and frameworks, including our adapted DINO. To fairly compare the networks, we picked networks that have approximately the same number of parameters.
Table~\ref{tab:result} presents the results for this comparison. DINO-CXR outperforms other methods in terms of accuracy, AUC, and F-1 score. Adapted DINO outperforms BYOL, SimCLR, and SimSaim by large margins, regardless of whether we use CNN or vision transformer as the backbone.
\renewcommand{\arraystretch}{1.4}
\begin{table}[!h]
    \centering
    \begin{tabular}{lcccccc}
        \Xhline{2\arrayrulewidth}
          Method & Network & Framework & Params (Million) & ACC & AUC & F1-score \\ \hline
          \rowcolor{LightCyan}
         DINO-CXR & ViTAEv2 & Adapted DINO & 19.35 & \textbf{95.66} & \textbf{0.9553} & \textbf{0.9613} \\ 
         & ResNet-50 & Adapted DINO & 23.5 & 94.10 & 0.9405 & 0.9466  \\ 
         & ViT-S/16 & Adapted DINO & 22.05 & 94.10 & 0.9437 & 0.9453  \\ 
         & ViTAEv2 & SimSiam & 19.35 & 83.01 & 0.8181  & 0.8587  \\ 
         & ViTAEv2 & SimCLR & 19.35 & 93.31 & 0.9322 & 0.9402  \\ 
         & ViTAEv2 & BYOL & 19.35 & 93.32 & 0.9179 & 0.9321  \\ \hline
         & ResNet-50 & SimSiam & 23.5 & 91.65 & 0.9160  & 0.9248  \\ 
         & ResNet-50 & SimCLR & 23.5 &  93.40 & 0.9329 & 0.9410  \\ 
         & ResNet-50 & BYOL & 23.5 & 91.65 & 0.9264 & 0.9130  \\ 
         & ViT-S/16 & SimSiam & 22.05 & 73.77 & 0.7064  & 0.8054  \\ 
         & ViT-S/16 & SimCLR & 22.05 & 89.56 & 0.8880 & 0.9106  \\ 
         & ViT-S/16 & BYOL & 22.05 & 92.24 & 0.9179 & 0.9321  \\ 
         \Xhline{2\arrayrulewidth}

    \end{tabular}
    \vspace{3px}
    \caption{Ablation study results on the Cell dataset. The performance of models was measured by accuracy (\%), area under the curve (AUC), and F1-score.}
    \label{tab:result}
\end{table}

\subsection{Pneumonia Detection}
Next, we evaluate the performance of DINO-CXR for pneumonia detection. Cell dataset is used as the target dataset to compare DINO-CXR with state-of-the-art self-supervised methods for pneumonia detection. Table~\ref{tab:compare with SOA} shows the results. DINO-CXR outperforms other methods in terms of accuracy and achieves comparable results in terms of AUC. We did not compare F-1 score as that score was not reported in other works.
\renewcommand{\arraystretch}{1.3}
\begin{table}[!h]
    \centering
    \begin{tabular}{cccc}
        \Xhline{2\arrayrulewidth}
          Dataset & Method & ACC & AUC \\ \hline
          \multirow{3}{2cm}{Cell dataset}& Gazda et. al~\cite{gazda2021self} &  91.5 & \textbf{97.7}\\
          & Kermany et. al \cite{kermany2018identifying} &  92.8 & 96.8\\
          \rowcolor{LightCyan}
          & DINO-CXR & \textbf{95.65} & 95.53 \\ \hline
          \multirow{3}{3cm}{COVIDGR dataset} & Gazda et .al\cite{gazda2021self} & 78.4 & 87.1 \\
          & COVID-SDNet\cite{tabik2020covidgr} & 76.18$\pm$2.70 & - 
          \\
          \rowcolor{LightCyan}
          & DINO-CXR & \textbf{76.47$\pm$3.53} & 75.78$\pm$4.22 \\
          
          \Xhline{2\arrayrulewidth} 

    \end{tabular}
    \vspace{3px}
    \caption{Comparing DINO-CXR with SOTA \textbf{SSL} methods. The numbers for other
methods are obtained from~\cite{panetta2021automated,gazda2021self}}
    \label{tab:compare with SOA}.
\end{table}

\subsection{COVID-19 Detection}

Finally, we evaluate DINO-CXR for COVID-19 detection compared to~\textit{supervised} and \textit{self-supervised} methods. The COVDIGR dataset is used for this purpose. The dataset is split into 90\% training, and 10\%
testing sets, 10\% of the training data as validation. The performance of DINO-CXR is assessed using the average and standard deviation values of the 5 different executions performed on the 5-fold cross-validation.
As one can see in Table~\ref{tab:COVID-result} and Table~\ref{tab:compare with SOA}, DINO-CXR outperforms other methods in terms of accuracy and achieves comparable results in other metrics while using significantly less labeled data. 
For instance, compared to COVIDNet-CXR~\cite{wang2020covid}, we use roughly 6\% (725 vs. 13,975) of the labeled data for fine-tuning.

\renewcommand{\arraystretch}{1.3}
\begin{table}[!h]
    \centering
    \resizebox{\columnwidth}{!}{%
    \begin{tabular}{p{4cm}c|cc|ccc}
        \Xhline{2\arrayrulewidth}
           & & \multicolumn{2}{c}{Non-COVID-19} & \multicolumn{3}{|c}{COVID-19} \\ \hline
          Method & ACC & F1-Score & Precision & F1-Score & Precision & Recall \\ \hline
          COVIDNet-CXR~\cite{wang2020covid} & 67.82$\pm$6.11 & 73.31$\pm$3.79&3.36$\pm$6.15 & 56.94$\pm$5.05 & 81.65$\pm$6.02 & 46.82$\pm$17.59 \\
          COVID-CAPS~\cite{afshar2020covid} & 65.34$\pm$3.26 & 65.15$\pm$5.02& 65.62$\pm$3.98 & 64.87$\pm$4.42 & 66.07$\pm$4.49 & 64.93$\pm$9.71 \\
          COVID-SDNet~\cite{tabik2020covidgr} & 76.18$\pm$2.70 & 76.94$\pm$2.82& 74.74$\pm$3.89 & 75.71$\pm$3.35 & 78.67$\pm$4.70 & 72.59$\pm$6.77 \\
          Panetta et. al~\cite{panetta2021automated} & 75.11$\pm$1.76 & 75.86$\pm$2.11& 74.75$\pm$3.61 & 74.02$\pm$3.15 & 76.41$\pm$7.38 & 72.65$\pm$6.83 \\
          \rowcolor{LightCyan}
          DINO-CXR & \textbf{76.47$\pm$3.53} & 78.03$\pm$1.96 &  73.49$\pm$5.5 & 72.86$\pm$7.13 & 79.93$\pm$1.94 & 66.93$\pm$11.72 \\
         \Xhline{2\arrayrulewidth}
    \end{tabular}
    }
    \vspace{3px}
    \caption{Comparative analysis of the proposed method with \textbf{supervised} deep learning based methods  on the COVIDGR dataset. The numbers for other methods are obtained from~\cite{panetta2021automated}.}
    \label{tab:COVID-result}
\end{table}

%% file: conclusion.tex
\section{Conclusion}
The combination of self-supervised pre-training and supervised fine-tuning has shown success in image recognition, particularly in scenarios where labeled samples are scarce. However, this approach has not been widely investigated in medical image analysis.\par
In this paper, we proposed DINO-CXR which is a novel adaptation of DINO based on a vision transformer for chest X-ray classification. Through extensive experiments, we showed the effectiveness of DINO-CXR and also demonstrated that DINO-CXR outperforms other methods for pneumonia and COVID-19 detection in terms of accuracy and achieves comparable results in terms of precision and F-1 score while requiring significantly less labeled data.

To the best of our knowledge, this is the first study to investigate the impact of different backbones on self-supervised pre-training approaches, and also show the benefit of inductive bias and its effectiveness on chest X-ray classification. We anticipate that this paper will contribute to the widespread adoption of self-supervised approaches in medical image analysis.